\documentclass[11pt,a4paper]{article}

\usepackage{amsmath}
\usepackage{amssymb}
\usepackage{amsthm}
\usepackage{color}
\usepackage{mathrsfs}
\usepackage[text={16truecm,23truecm}]{geometry}
\usepackage{graphicx}
\usepackage[OT4]{fontenc}
\newtheorem{propo}{Proposition}
\newcommand{\Farias}[1]{} 
\newcommand{\Enk}[1]{}
\def\openone{\leavevmode\hbox{\small1 \normalsize \kern-.64em1}}

\title{Static Quantum Games Revisited}
\author{Marcin Markiewicz
\footnote{ \noindent
Institute of Theoretical Physics and Astrophysics, 
University of Gda\'nsk,
Wita Stwosza 57,
80952 Gda\'nsk, Poland.
}
\and Adrian Kosowski
\footnote{ \noindent
Department of Algorithms and System Modeling,
Gda\'nsk University of
Technology,
Narutowicza 11/12,
80233 Gda\'nsk, Poland.
}
\and Tomasz Tylec
\footnotemark[1]
\and Jaros\l{}aw Pykacz 
\footnote{ \noindent
Institute of Mathematics,
University of Gda\'nsk,
Wita Stwosza 57,
80952 Gda\'nsk, Poland.
}
\and Cyril Gavoille
\footnote{ \noindent
LaBRI - University of Bordeaux,
351 cours de la Lib\'eration,
33405 Talence cedex, France.
}
}

\date{}

\begin{document}

\maketitle

\begin{abstract}
The so called \emph{quantum game theory} has recently been proclaimed as one of the new branches in the development of both quantum information theory and game theory. However, the notion of a quantum game itself has never been strictly defined, which has led to a lot of conceptual confusion among different authors. In this paper we introduce a new conceptual framework of a \emph{scenario} and an \emph{implementation} of a game. It is shown that the procedures of ``quantization'' of games proposed in the literature lead in fact to several different games which can be defined within the same scenario, but apart from this they may have nothing in common with the original game. Within the framework we put forward, a lot of conceptual misunderstandings that have arisen around ``quantum games'' can be stated clearly and resolved uniquely. In particular, the proclaimed essential role of entanglement in several static ``quantum games'', and their connection with Bell inequalities, is disproved.
\end{abstract}

\section{Introduction}
\label{s1}

Since the late 1990's, numerous works on so-called ``quantum games'' have appeared (e.g. \cite{EWL99, BH01, LJ03, CWLW04, FSSLH09, AV08a, AV08b}), resulting sometimes in a redefinition of basic notions of a game-theoretic nature, not always in a precise way. The goal of this note is to clarify the terminology and concepts used by different authors.

The abstract concept of a static \emph{game} is defined through a set of players, the set of their strategies, and a set of payoff functions (cf.~e.g.~\cite{OR94} for a formal introduction to the topic).
This notion is sufficient for most considerations of a game-theoretical nature, e.g.~finding Nash equilibria. However, the very idea of a game presupposes the existence of real-world situations in which players undertake various actions in accordance with their chosen strategies, and they obtain payoffs according to the achieved final state of a game. Any such real-world situation can be called \emph{implementation} of a game. In particular, an implementation of a game can be understood as a physical system where strategies are represented by some transformations on the system, and payoffs are represented by results of appropriate measurement performed on the system. Obviously, one abstract game may have many different implementations, which does not affect the essential properties of the game.

Throughout this note, we introduce and use the concept of an $n \times k$ \emph{scenario}, which intuitively may be
understood as a framework of inputs/outputs, rather like a computer
science problem (cf.~\cite{Wing06}).
The notion of a \emph{scenario} in the context of game theory is a new concept, and is defined with the aim of clarifying the way in which the quantum framework is introduced into game theory.
Scenarios are given as triples $(\mathcal X, \mathcal Y,\$)$, where $\mathcal X, \mathcal Y \subseteq \{0,\ldots,k-1\}^n$ are valid inputs and outputs for the scenario, respectively, while $\$ = (\$_1, \ldots, \$_n)$, where each $\$_i \colon \mathcal X \times \mathcal Y \to \mathbb{R}$, is an evaluating function. Some examples of $n\times 2$ scenarios are based on well-known problems in game theory and distributed computing, including:
\begin{itemize}
\item The \emph{minority scenario}, which takes no input $(\mathcal X = \emptyset)$, allows single-bit output for each player $(\mathcal Y = \{0,1\}^n)$, and for a bit vector $(y_1,y_2,\ldots,y_n)\in \mathcal Y$ we have $\$ (y_1,y_2,\ldots,y_n) = (c_1,c_2,\ldots,c_n)$, where $c_i = 1$ if $| \{y_j : y_i=y_j, 1\leq j \leq n\}| < n/2$, where $|A|$ denotes cardinality of a set $A$, and $c_i=0$ otherwise.
\item The \emph{Prisoner's Dilemma scenario} for $n=2$, which takes no input $(\mathcal X = \emptyset)$, allows single-bit output for each player $(\mathcal Y = \{0,1\}^2)$, and an evaluating function such that $\$_1(1,0)=\$_2(0,1) > \$_1(0,0)=\$_2(0,0) > \$_1(1,1)=\$_2(1,1) > \$_1(0,1)=\$_2(1,0)$, and $\$_1(0,0)\geq \left(\$_1(1,0)+\$_1(0,1)\right)/2$.
\item The \emph{Battle of the Sexes scenario}, which is similar to that of the Prisoner's Dilemma, except for different evaluating functions: $\$_1(0,0)=\$_2(1,1) > \$_1(1,1)=\$_2(0,0) > \$_1(0,1)=\$_2(0,1) = \$_1(1,0)=\$_2(1,0)$.
\item The \emph{modulo-4 scenario}~\cite{ZB09} for $n=3$, where input $\mathcal X = \{(x_1,x_2,x_3) : x_1, x_2, x_3 \in \{0,1\} \text{\ and\ } x_1+x_2+x_3\equiv 0 \mod 2\}$, output $\mathcal Y = \{0,1\}^3$, and $\$(y_1, y_2, y_3) = (1,1,1)$ if $2(y_1+y_2+y_3) \equiv x_1+x_2+x_3 \mod 4$, and $\$(y_1, y_2, y_3) = (0,0,0)$, otherwise.
\end{itemize}
All of these scenarios are inspired by some $n\times2$ games\footnote{That is games with $n$ players, each equipped with two pure strategies.}, but in the context of this discussion and prior works, may in fact lead to games different from the original ones. The first three games (Minority Game, Prisoner's Dilemma and The Battle of the Sexes) are well known and do not require additional comments. The modulo-4 game is a sequential game. In the first stage, one player gives one bit to each of his three opponents, in such a way that the sum of distributed bits equals $0$ or $2$. In the second stage, each of the opponents produces bit output, and no communication between the players is allowed. If the doubled sum of the outputs is congruent to the sum of inputs modulo $4$, the first player gives payoff $0$, whereas his opponents receive $1$. Otherwise, the payoff of the first player is $1$, whereas his opponents receive $0$.

Within a single scenario, we can define different \emph{procedures} for transforming input into output. This sort of procedure is a general concept and need not be described by any computable function. Here, we confine ourselves to the so called \emph{EWL-type procedure}, inspired by the original model of  Eisert, Wilkens and Lewenstein (EWL)~\cite{EWL99}.
Although only $n\times 2$ scenarios  can be furnished with the EWL-type procedure, this notion is sufficient for discussing most cases which appear in the literature concerning ``quantum'' versions of games.

The EWL-type procedure is parameterized by the five-tuple $P=(n,H,J,\rho, \mathfrak{U})$, where $n$ is the number of players, $H, J$ are unitary operators on $(\mathbb{C}^2)^{\otimes n}$, $\rho$ is some initial $n$-qubit state, and $\mathfrak{U}$ is a compact subset of $\mathsf{SU}(2)$\footnote{$\mathsf{SU}$ denotes \emph{the special unitary group}.}.
The procedure works as follows: the initial $n$-qubit state $\rho$ undergoes evolution under the successive actions of operator $J$, of some operator $\mathcal U = U_{1,x_1}\otimes\dots\otimes U_{n,x_n}$, where $U_{i,x_i} \in \mathfrak{U}$ are operators (encoding action of $i$-th player) dependent on input $X=(x_1,\ldots, x_n)\in \mathcal X$, and finally of operator $H$. The final state $\rho_f$ is measured in the standard basis, and the measurement result $Y\in\{0,1\}$ is treated as the output of the system. Formally, the probability $p(Y|X)$ of obtaining output $Y\in \mathcal Y$ for input $X$ under procedure $P$ is given as follows:
\begin{align}
\label{pr1}
p(Y|X)\equiv \operatorname{Pr}[Y&=(y_1,\ldots,y_n)\ |\ U_{1,x_1},\ldots,U_{n,x_n}] = \operatorname{Tr} (Q_{y_1} \otimes \dots \otimes Q_{y_n} \rho_f)\\
\intertext{where:}
	\rho_f &= H \mathcal U J \rho (H \mathcal U J)^\dagger,\nonumber\\
	Q_0 &= |0\rangle\langle0| \text{ and } Q_1 = |1\rangle\langle1|,\nonumber\\
	\mathcal U &= U_{1,x_1}\otimes\dots\otimes U_{n,x_n}.\nonumber
\end{align}
The most important property of the described EWL-type procedure is formulated in the following proposition.

\begin{propo}
\label{prop1}
Assume that some $n\times2$ scenario with no input, $F = (\emptyset, \mathcal Y, \$)$, is furnished with some EWL-type procedure $P$, where the actions of the players are defined through their choices of local unitary operators $U_1,\ldots,U_n$. Then procedure $P$ within scenario $F$ is an implementation of some static (possibly continuous) $n$-player game, which has at least one Nash equilibrium.
\end{propo}
\begin{proof}
Indeed, procedure $P$ implements a game which has $n$ players, a set of pure strategies corresponding to some parametrization $(\theta, \varphi, \chi)$ of elements of $\mathfrak{U}\subseteq \mathsf{SU}(2)$, and a payoff function $\yen = (\yen_1, \ldots,\yen_n)$, given for player $i$ as:
\begin{equation}
\label{pay1}
\yen_i ((\theta_1, \varphi_1, \chi_1),\ldots,(\theta_n, \varphi_n, \chi_n)) =
\sum_{y_1,\ldots,y_n \in \{0,1\}}
	\mathcal \$_i (y_1,\ldots,y_n) \operatorname{Tr} (Q_{y_1} \otimes \dots \otimes Q_{y_n} \rho_f).
\end{equation}
In the case of discrete sets $\mathfrak{U}$ every such game as $n$-player game with a finite set of strategies  admits Nash equilibrium (at least in mixed strategies) according to Nash Theorem \cite{N50}.
In the case of continuous sets $\mathfrak{U}$ defined game as a continuous game with compact set of strategies and continuous payoff functions, admits a Nash equilibrium (at least in mixed strategies corresponding to probability measures on the set of strategies) according to Glicksberg's Theorem~\cite{G52}.
\end{proof}

In the case of a general $n\times k$ scenario (that is a scenario motivated by some $n$-player game, where each player has $k$ pure strategies), the above defined EWL-type procedure can be  straightforwardly generalized, using states and operators on $k$-dimensional Hilbert space, instead of $2$-dimensional Hilbert space of qubits.  Then, a similar proposition can be formulated. We do not formulate this general version of the proposition for the sake of clarity of further considerations.

\section{How much quantumness is there in static quantum games?}

\subsection{Two-player ``quantum games''}

In some comments~\cite{EP02,LJ03,P07}, it is pointed out that by redefining a classical implementation of certain games in a quantum setting, one obtains an implementation of a different game, having a wider set of strategies. This discussion may be more precisely restated within the introduced framework, using the notions of a game, an implementation, and a scenario. Indeed, each of the scenarios defined in the Introduction (the minority scenario, the Prisoner's Dilemma scenario, etc.) can be furnished with a class of procedures, some of which constitute implementations of different games, with the original game being the motivation for defining a given scenario. For example, within the Battle of the Sexes scenario we can define several EWL-type procedures, corresponding to the implementation of the standard Battle of the Sexes game and to some different versions of the ``quantum Battle of the Sexes game'' (cf.~\cite{DXLZH01,EWL99,JJN04,F09,MW00}):
\begin{itemize}
	\item $P_1=(n=2, J=\openone, H=\openone, \rho=|00\rangle\langle 00|, \mathfrak U=\{\sigma_x, \openone\})$
	\item $P_2=(n=2, J=\openone, H=\openone, \rho=\frac{1}{2}(|00\rangle\langle 00|+|00\rangle\langle 11|+|11\rangle\langle 00|+|11\rangle\langle 11|), \mathfrak U=\{\sigma_x, \openone\})$
	\item $P_3=(n=2, J=\openone, H=\openone, \rho= \frac{1}{2}\bigl(1 - (\epsilon_{1} +
\epsilon_{2})\bigr)\bigl(|00\rangle\langle00| +
|11\rangle\langle11|\bigr) + \epsilon_{1}|01\rangle\langle01| +
\epsilon_{2}|10\rangle\langle10|, \mathfrak U=\{\sigma_x, \openone\})$, where $|\epsilon_{1} -
\epsilon_{2}| > 0$
	\item $P_4=(n=2, J=e^{i\gamma \sigma_y \otimes \sigma_y}, H=J^{\dagger},\rho=|00\rangle\langle 00|, \mathfrak U=\{U(\theta, \phi)\}\subset \mathsf{SU(2)})$
	\item $P_5=(n=2, J=e^{i\gamma \sigma_y \otimes \sigma_y}, H=J^{\dagger},\rho=|00\rangle\langle 00|, \mathfrak U=\{U(\theta, \phi, \chi)\}= \mathsf{SU(2)})$
\end{itemize}
$P_1$ is obviously an implementation of the standard Battle of the Sexes game, while $P_2$, $P_3$, $P_4$ and $P_5$  are implementations of the so called  ``quantum Battle of the Sexes'' game in the versions from \cite{MW00}, \cite{F09}, \cite{EWL99}, and \cite{DXLZH01} respectively.
$P_1$, $P_2$ and $P_3$ are implementations of three different discrete games, which have identical sets of strategies, but different payoff functions, while $P_4$ and $P_5$ are implementations of two different continuous games, which have different sets of strategies, different payoff functions and different Nash equilibria~\cite{JJN04}. Consequently, games implemented by $P_2$, $P_3$, $P_4$ and $P_5$ have nothing in common with the standard Battle of the Sexes game except for the fact that all these four games can be introduced within the same scenario, motivated by the standard Battle of the Sexes game.

Exactly the same conclusions are reached when considering ``quantum versions'' of  other static games with $n$ players equipped with  two strategies (e.g. the ``quantum Prisoner's Dilemma'' studied in~\cite{EWL99}). In each case the ``quantization'' of a game is just defining a different (usually continuous) game within a scenario motivated by the original game. The only common element of the original and modified games is the same number of players, and the fact that the payoff functions from the original game enter Eq.~(\ref{pay1})  for the payoff functions of the ``quantized'' version.

Although we have only discussed ``quantization'' of games with two strategies for each player, analogous reasoning can be performed for more general situation using $n\times k$ scenarios furnished with appropriate procedures, leading to similar conclusions.

\subsection{When does entanglement help?}
\label{s2}

Several papers~\cite{CWLW04,FSSLH09,BH01} have considered quantum versions of \emph{minority games} in which an entangled state, initially shared by all parties, and finally measured in some basis, leads to new Nash equilibria, increasing the payoff of all players with respect to the classical versions.\footnote{For such Nash equilibria to appear, some authors~\cite{FSSLH09} assume that the strategies used by all players are identical. This assumption is logically flawed in a game-theoretic context. However, for an appropriately chosen initial state, new Nash equilibria appear regardless of this assumption.} Using the framework introduced herein we can easily show that within the \emph{minority scenario} (or in fact, more generally, all scenarios with empty input), this type of effect can always be achieved without resorting to entanglement, using purely classical resources.

Indeed, consider three different EWL-type procedures within the \emph{4-player minority scenario}:
\begin{itemize}
	\item $P_1=(n=4, J=\openone, H=\openone,\rho=|0000\rangle\langle 0000|, \mathfrak U=\{\sigma_x, \openone\})$
	\item $P_2=(n=4, J=\openone, H=\openone,\rho_{in}, \mathfrak U= \{\sigma_x, \openone\})$, where
	\begin{align*}
\rho_{in} & = \tfrac{1}{8} (
Q_0\!\otimes\! Q_0\!\otimes\! Q_0\!\otimes\! Q_1+
Q_0\!\otimes\! Q_0\!\otimes\! Q_1\!\otimes\! Q_0+
Q_0\!\otimes\! Q_1\!\otimes\! Q_0\!\otimes\! Q_0+
Q_1\!\otimes\! Q_0\!\otimes\! Q_0\!\otimes\! Q_0+ \nonumber\\
&+ Q_1\!\otimes\! Q_1\!\otimes\! Q_1\!\otimes\! Q_0+
Q_1\!\otimes\! Q_1\!\otimes\! Q_0\!\otimes\! Q_1+
Q_1\!\otimes\! Q_0\!\otimes\! Q_1\!\otimes\! Q_1+
Q_0\!\otimes\! Q_1\!\otimes\! Q_1\!\otimes\! Q_1)
\end{align*}
	\item $P_3=(n=4, J=\openone, H=\openone, \varrho_{in}=|\psi_{in}\rangle\langle \psi_{in}|, \mathfrak U=\{U(\theta, \phi, \chi)\}= \mathsf{SU(2)})$, where:
	\begin{equation*}	 |\psi_{in}\rangle=\frac{\alpha}{\sqrt{2}}\left(|0000\rangle+|1111\rangle\right)+\frac{\sqrt{1-\alpha^2}}{2}\left(|01\rangle+|10\rangle\right)^{\otimes 2}
	\end{equation*}
\end{itemize}
Here, $P_1$ corresponds to the classical minority game, $P_3$ is the ``quantum minority game'' from~\cite{FSSLH09}, and $P_2$ is some procedure in which the initial state $\rho_{in}$ is separable and no entanglement ever appears (since $J=H=\openone$).

Within the $4$-player minority scenario, let $p(Y)$ be the probability of obtaining output  $ Y = (y_1, y_2, y_3, y_4) $, $y_i \in \{0,1\}$, as given by Eq.~(\ref{pr1}) of Section 1.  Then, there exists a mixed separable state $\rho = \sum_{Y\in \{0,1\}^4} \left[p(Y) (Q_{y_1}\otimes Q_{y_2}\otimes Q_{y_3}\otimes Q_{y_4})\right]$, such that the trivial strategy $U_i = \openone$ results in the same output distribution $\{(Y,p(Y))\}$. Thus, in particular, the considered procedure $P_2$ includes a Pareto-optimal Nash equilibrium, corresponding to $p(Y)=1/8$ for bit-outputs $Y$ which have an odd number of ones, and $p(Y)=0$ otherwise. This strategy leads to exactly the same payoff of $1/4$ as an analogous strategy which appears when considering procedure $P_3$~\cite{FSSLH09}.  In this sense, the difference between the two games implemented by procedures $P_1$ and $P_3$ is the same as in the case of procedures $P_1$ and $P_2$, and so no quantum entanglement is required to achieve such a distinction.

We remark that in the procedure $P_2$, the separable state $\rho_{in}$ may be considered in a purely classical setting as follows. Initially, a helper randomly picks one of two decks of 4 cards each, one with numbers $\{0,0,0,1\}$ written on the hidden faces, and the other with numbers $\{1,1,1,0\}$. Then, the helper shuffles the cards and randomly gives them out to the 4 players of the game. At this point, the only Pareto-optimal strategy for each player is to read out the value of the card received, achieving an expected payoff of $1/4$ for each player.

Whereas the considered effect never requires entanglement for scenarios with no input, entanglement may sometimes be beneficial in the case of scenarios with non-trivial input. This is the case, for example, for the modulo-4  scenario. The analysis follows from that used in the discussion of quantum distributed computing models~\cite{GKM09}.
In order to introduce the concept of an implementation of a game for scenarios with non-trivial input we have to reformulate conclusions from the Proposition \ref{prop1}. Let $P$ be some EWL-type procedure for some $n\times 2$  scenario  $F = (\mathcal X, \mathcal Y, \$)$, where elements  $X\in\mathcal X$ and $Y\in\mathcal Y$ are given by vectors $X=(x_1,\ldots, x_n),\;Y=(y_1,\ldots, y_n)$. The implementation of a game is then understood in terms of repeated iterations of the game process, where all inputs are chosen with equal probability, and payoffs are averaged over the statistical sample. Formally, procedure $P$ implements a game which has $n$ players, a set of pure strategies for $i$-th player corresponding to some parametrization $\{(\theta_0, \varphi_0, \chi_0),(\theta_1, \varphi_1, \chi_1)\}$ of two elements $\{U_{i,0}, U_{i,1}\} $ of $\mathfrak{U}\subseteq \mathsf{SU}(2)$ (corresponding to two different values of possible input $x_i$ for $i$-th player), and a payoff function $\yen = (\yen_1, \ldots,\yen_n)$, given for player $i$ as:
\begin{equation}
\label{pay2}
\yen_i  = \frac{1}{|\mathcal X|}\sum_{x_1,\ldots,x_n \in \{0,1\}}
\sum_{y_1,\ldots,y_n \in \{0,1\}}
	\mathcal \$_i (x_1,\ldots,x_n,\ y_1,\ldots,y_n) \operatorname{Tr} (Q_{y_1} \otimes \dots \otimes Q_{y_n} \rho_f(x_1,\ldots, x_n)).
\end{equation}
Let us define the following EWL-type procedure within the modulo-4 scenario:
\begin{equation}
P_{GHZ}=(n=3, J=\openone, H=\openone, \rho=|GHZ\rangle\langle GHZ|, \mathfrak U=\mathsf{SU(2)})
\label{ewlghz}
\end{equation}
where $|GHZ\rangle=\frac{1}{\sqrt{2}}\left(|000\rangle+|111\rangle\right)$. According to former considerations, this procedure is an implementation of a $3$-player continuous game. The important fact is that this game has strategy profiles, which maximize payoffs for all players in the sense of Eq. \eqref{pay2}. It can be proven, that this situation within modulo-4 scenario cannot be achieved by implementations using separable states only \cite{GKM09}.

To sum up, the reason why entanglement is essential for the modulo-4  scenario is that among all games that can be implemented by procedures within this scenario, only those using entangled states in implementations have strategy profiles which lead to maximal possible payoff for each player. Nevertheless, all those games are still different static games, and all considerations of the beneficence of entanglement can be reasonably performed at the level of scenarios, and not of the games themselves. It should be emphasized that the requirement of entanglement in such cases is a consequence of the specific definition of the \emph{scenario}, and does not imply any essential quantumness of the underlying \emph{games} themselves.

\subsection{Notation in quantum games vs. Bell-type inequalities}
\label{s3}
Some authors have recently remarked on the apparent similarities between non-cooperative quantum minority games and Bell inequalities \cite{FSSLH09}. In view of the previous sections, effects related to entanglement are not observed in the minority scenario, hence such a similarity must be illusory. We elaborate this point below.

Since game theory is based on probabilistic framework (in the sense of Kolmogorovian probability theory), the discussion of any possible relation to Bell-type inequalities should treat them in purely probabilistic terms, not referring to philosophical assumptions behind them \cite{ZB09}. Bell-type inequalities introduce bounds on values of correlation functions of random variables in Kolmogorovian probability theory. From the theorem of Gelfand and Naimark \cite{STR00} we conclude that if the set of observables of a system can be represented as Abelian algebra, then all the observables from this algebra may be treated as random variables on some common probability space. However, Quantum Mechanics introduces non-Abelian algebras of observables, which cannot be represented as random variables on a common probability space, consequently leading to the violation of Bell-type inequalities.
From this point of view we can say that Quantum Mechanics is a generalization of classical probability. Each compatible set of observables can be individually treated classically (as random variables), but the whole algebra of observables obviously cannot. This fact leads to the concept of \emph{contextuality}: each measurement context (that is the choice of compatible observables) defines a different classical statistical model.

After this short introduction the improper manner of applying Bell-type inequalities to given experimental situation in ``quantum minority game'' from \cite{FSSLH09} (defined by procedure $P_3$ in section \ref{s2}) is clearly revealed. Bell-type inequalities involve correlations of measurement results on \emph{different observables chosen independently by each observer}, and this framework is the only reasonable one within which we can discuss correlations of independent local measurements. If violation of Bell-type inequalities is achieved, then this implies that the Bell-type inequality involves mean values of random variables (assigned to quantum observables) coming from different, incompatible classical probabilistic models (different \emph{experimental contexts} induced by independent choices of measurement settings applied by each observer).

However, the expression for the payoff function in \cite{FSSLH09} is based on the relative frequencies of measurement outcomes (i.e., different real eigenvalues) in measurements of \emph{a single global observable}, which of course does not fit into the discussed scheme of applying Bell-type inequalities to an experimental situation. One may argue that local unitary operations performed by each of the players define a measurement basis for the final measurement, hence approaching the correlational scheme of Bell-type inequalities. This argument is not valid in this discussion: according to the rules imposed on players in \cite{FSSLH09}, the performed unitary operations are identical for each player, thus they define only one experimental context (i.e., the same local measurements). When there is only one experimental context, no Bell-type inequalities can be violated. Hence, the violation of Bell-type inequality in the experimental situation presented in the ``quantum minority game'' from~\cite{FSSLH09} is of no meaning.

The misunderstanding arises because of the combinatorial similarity between the expression for the payoff function and MABK Bell polynomial (\cite{M90, A92, BK93}). However, if the payoff function reveals any similarity to the MABK Bell polynomial, the only reasonable experimental situation to which it would refer is a setup with 4 observers, each measuring two different observables. But this is not the situation in ``quantum minority game''. There is no similarity in any other sense that combinatorial between the MABK polynomial and the payoff function. We can generalize this to the obvious statement that whenever any function referring to a physical quantity is similar in form to expressions forming Bell-type inequalities, it does not mean that we can reasonably discuss for this quantity the violation of the Bell-type inequality, which is applicable to only one experimental situation.

\section{Conclusions}

The formalization of quantum procedures in static games, which we propose herein, reveals two natural paths of future research. On the one hand, it is interesting to study scenarios with non-trivial input, which may potentially reveal genuinely quantum effects. Such effects, including any possible relations to Bell inequalities, can never be observed for scenarios with zero input. On the other hand, one may potentially attempt to introduce the concept of a \emph{quantum game}, which relies on a quantization of the underlying mathematical formalism of a game, rather than modifications to scenarios.

\section{Acknowledgements}
Marcin Markiewicz gratefully acknowledges the financial support of the Q-ESSENCE 7FP (Grant agreement number 248095).
Jaroslaw Pykacz gratefully acknowledges the financial support of University of Gdansk grant BW/5100-5-0156-9.

\bibliographystyle{abbrv}

\end{document}